# THE PADME TRACKING SYSTEM[*]

G. Georgiev[1][**], V. Kozhuharov, L. Tsankov

[1]Faculty of Physics, University of Sofia "St. Kl. Ohridski", Sofia, Bulgaria

*Abstract: The Positron Annihilation into Dark Matter Experiment (PADME) at LNF-INFN Linac aims to perform a search for dark photons in positron-on-target annihilation process. A key component of the setup is the tracking system which allows vetoing the bremsstrahlung-induced background. Different solutions for the detector will be shown and will be discussed. Attention will be paid to the possibility to construct a hybrid tracker based on plastic scintillator fibers read out by CCD matrices.*

*Keywords: Dark photon, tracker, hodoscope, CCD camera*

## 1. Introduction

The question about the nature of Dark Matter is still unresolved in Particle Physics. While there are robust indications about its existence, the way it interacts and what it is made of remains unclear. Recently, a number of experimental activities were initiated to probe the possible extension of the Standard Model. A large family of models predicts the existence of a new force describing the interactions among the dark matter particles [1]. The gauge boson, A', or the so-called dark photon of this new force, could also act as a mediator between the Standard Model particles and the hidden dark matter sector. The coupling of A' to the visible fermions may be very small and therefore it might have escaped yet unobserved.

## 2. PADME at DAΦNE Linac

The Positron Annihilation into Dark Matter Experiment (PADME) [2,3] at the Frascati Beam Test Facility (BTF) plans to address the existence of the dark photon by studying the positron-on-target annihilation process – $e^+e^- \rightarrow A'\gamma$. If a dark photon in the accessible mass range exists, it will appear as a peak in the missing mass distribution $M^2_{miss}=(P_{e+}+P_{e-}-P_\gamma)^2$, where $P_i$ are the four momenta of the particles. The major background comes from the two-photon annihilation process, three-photon annihilation, and bremsstrahlung from the beam positrons.

Schematic of experimental setup is shown in Fig. 1.

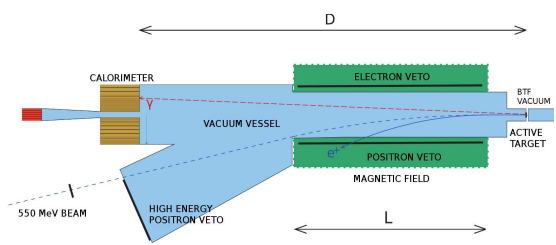

Figure 1. Schematic of the PADME experiment

The beam will be delivered by the DAΦNE Linac which is able to provide positrons with energy up to 550MeV, with an energy spread better than 1% [4,5]. A single bunch in PADME configuration will contain $5\times10^3$ positrons. The Linac is able to deliver up to 50 bunches per second, one of which is diverged to a beam monitor for better energy control and up to 49 are available at the BTF. Recent studies indicated that the length of the bunch could be increased up to 40ns without the degradation of the performance [6]. The major subdetectors of the PADME experiment are listed below:

**Active target: 100μm** thin detector made of polycrystalline diamond strips will serve as a target of the experiment. The strips will be in X and Y direction with the width of 1mm. The target will provide information about the beam position and profile for each bunch from the Linac.

**Calorimeter**: An inorganic crystal calorimeter made of BGO is planned for the measurement of the energy and the impact point of the recoil photon. It should provide a resolution better than 5% for $E_\gamma$=100MeV and position resolution better than 3mm. These quantities, together with the beam position at the active target, will determine $P_\gamma$ and are crucial for the reconstruction of the missing mass. The calorimeter will also serve as a veto for two- and three-photon annihilation events.

**Spectrometer**: Positrons that emit bremsstrahlung photons in the target will have lower energy and will be detected by a set of detectors for charged particles inside the magnetic field, as seen in Fig.2. Due to the small beam divergence, the correlation between the longitudinal impact point and the momentum provides ≈1% momentum resolution. This information will be combined with the data from the calorimeter to suppress the background from bremsstrahlung events, since the probability of a positron to emitting a photon with energy >1MeV in the diamond target is 5 orders of magnitude higher than for two-photon annihilation. Another detector on the side of the

---

deflection of the electrons will join the detector on the positron side and will help to keep events in which the dark photon had decayed, since they possess a genuine low-energy positron and would had been vetoed. A third detector, further along the deflected positron beam is planned, which will provide additional rejection of high-energy positrons that will undergo bremsstrahlung emission. The average occupancy of the spectrometer on the positron deflection side is ≈5 particles per bunch, over a total length of 1 m. The track matching will be based on the time difference between the track time in the spectrometer and the time of the electromagnetic shower in the calorimeter. That is why time resolution of the order of 500ps is necessary.

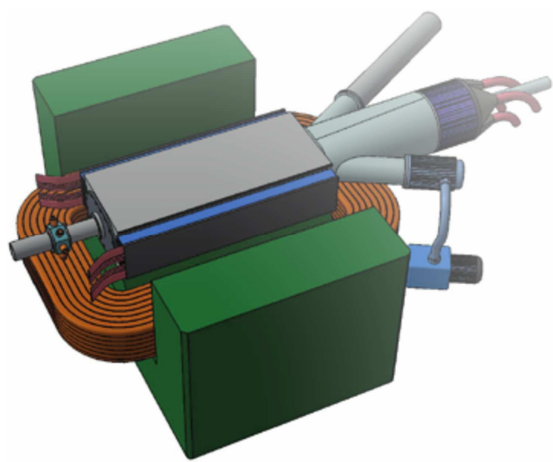

Figure 2. Schematic of the magnet of the PADME experiment. The charged particle detectors are placed at both sides of the vacuum chamber, parallel to the magnetic field.

With the presented setup and detectors, satisfying the requirements, a sensitivity down to $10^{-6}$ for the relative interactions strength with respect to the electrodynamics is achievable for A' mass in the interval ≈1÷24MeV. During the experiment R&D, several options were considered for the technology of the spectrometer detector of the PADME experiment. Two of them are listed below.

## 3. HYBRID SCINTILLATING FIBER DETECTOR

The spectrometer should have both good time resolution and good position resolution, translated into good momentum resolution. The latter requires the construction of a segmented detector with many channels. Such a detector could be made of layers of scintillating fibers. A 1m long detector could be built with 1000 fibers with 1mm diameter. However, the requirement to reconstruct precisely the time of the hits in every single fiber might lead to the necessity for a huge amount of fast photodetectors. This can be avoided by splitting the measurement of the two quantities – the position and the time – due to the possibility of registering the scintillation light on both fiber ends. The position can be determined by a slow, position-sensitive detector on one side, while the time can be measured by a fast photodetector attached to a group of fibers on the other, as shown in Fig. 3.

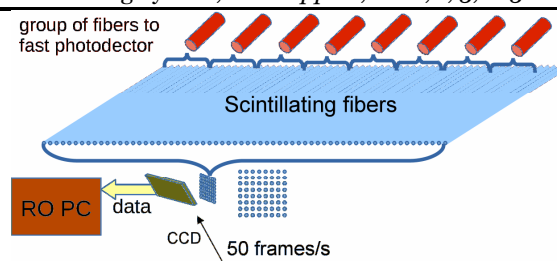

Figure 3. Schematic of a hybrid tracker made of scintillating fibers.

The bunch rate of 50Hz is slow enough to allow the application of a conventional CCD camera, able to provide information for each single fiber. This could help in reducing the number of necessary photodetectors, power supply and readout channels by a factor of 10.

### 3.1. Study of the CCD as a scintillation light detector

The fiber solution for the PADME charged particle detector was considered very attractive because of the precise position resolution it provided. However, the possibility of detecting scintillation light with a CCD had to be validated. In addition, the amount of light generated by the passage of a single particle through a fiber with 1mm diameter amounted to ≈100 photons at each fiber end.

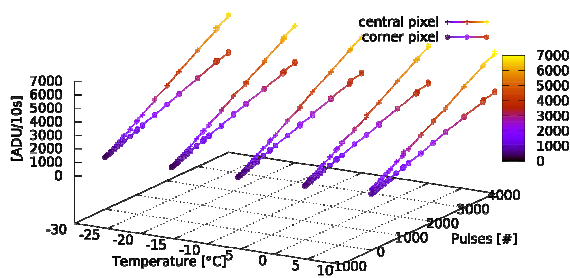

Figure 4. Response of the CCD to pulsed light from the LED driver

The CCD response was tested with a custom developed LED driver [7]. A high-performance camera for astronomical observations, SBIG 11000M, was exposed to a diffuse pulsed light. This full frame CCD camera featured 4008(H)×2672(V) pixels with 9μm pixels size, >40% quantum efficiency for 400nm light and Peltier element able to cool the matrix down to 30 degrees below the ambient temperature. This has a positive impact on the signal to noise ratio.

The response of two different pixels, one close to the center and one to the edge of the matrix, is shown in Fig. 4, for a different number of low-light-intensity pulses from the LED-driver. The chip temperature was varied, from -30 to +10 degrees Celsius. Both the edge and the central pixel provided the linear response for all the temperature conditions.

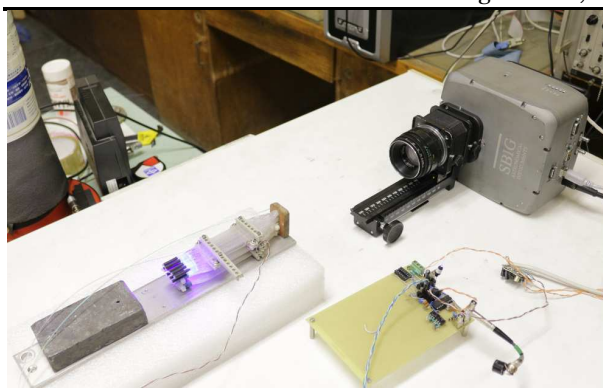

Figure 5. A photo of the SBIG 11000M CCD camera coupled to the focal length matching system, the LED-driver prototype, and the fiber tracker prototype

A small prototype of Kuraray SCSF-81 scintillating fibers was constructed, consisting of two planes of double-layer scintillating fibers and two planes with a single layer of scintillating fibers. The SBIG camera was coupled to the prototype through a focal length matching system, as seen in Fig. 5.

A $^{90}$Sr/$^{90}$Y radioactive source was placed on the upper surface of the tracker prototype. Data were collected at different temperatures and different exposure time. An image recorded with 1000s exposure time is shown in Fig. 6.

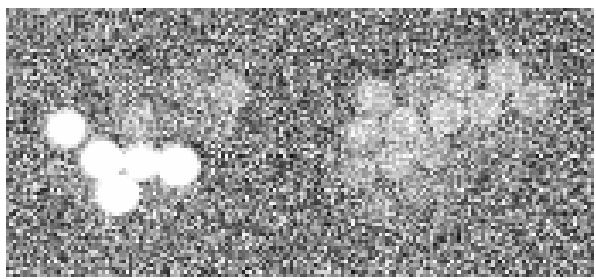

Figure 6. Image of a prototype of a fiber tracker exposed to beta rays from $^{90}$Sr/$^{90}$Y radioactive source.

The optical system was found to decrease the light intensity of the fiber image on the CCD matrix with respect to the emitted light by a factor as high as 100. However, the fiber pattern is clearly visible. While the qualitative result is that the CCD camera is a perspective detector for scintillation light due to ionizing radiation, the quantitative estimation of the minimal detectable absorbed dose (or fluence of high energy ionizing particles) is still ongoing.

4. SCINTILLATING STRIPS DETECTOR

All three charged particle detectors could be constructed from extruded plastic scintillator bars with fast photodetector readout, as shown in Fig. 7.

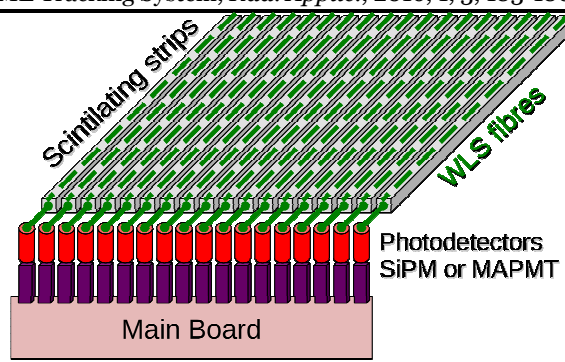

Figure 7. Schematic of a tracker made of scintillating strips

The light can be readout either by a SiPM directly coupled to the bar or through a WLS fiber placed in a groove along the bar. The WLS fiber will provide better signal uniformity and could allow detaching the photodetector from the active detector region. Currently, bars with square cross section of 1×1cm$^2$ are considered, with length determined by the gap between the magnet poles - <18cm.

A diffuse white reflective coating could be applied to the bars for better light collection. Since the thickness of the coating is ≈250μm, special care has to be taken to avoid the possibility of a particle passing through dead detector region. This can be done by rotating the individual bars at an angle of about 100mrad, as shown in Fig. 8.

The energy deposited in a plastic scintillator traversed by positrons with energy in the range from 50 MeV – 500 MeV was modelled with PENELOPE [8] and is shown in Fig. 9, for scintillator thickness of 5 mm and 10 mm. Since the particles are highly energetic, the individual distributions overlap. To estimate the contribution of the scintillator to the inefficiency of the detector, a tentative cut on the minimal detectable energy of 500 keV was assumed. It corresponds to O(10) photoelectrons on a photodetector with O(10%) quantum efficiency. Both, the 5 mm and the 10 mm solution provide of the order or less than 1% inefficiency in the whole positron energy range, as seen from Fig. 10.

*4.1. Photodetector choice*

Two major photodetector technologies are under consideration - multi anode photomultiplier (MAPMT) and silicon photomultiplier (SiPM). The SiPM has the advantage of lower operating voltage and that the sensor is relatively cheap. However, a more complicated power supply system with temperature stabilization and transimpedance preamplifier is necessary for their proper operation. On the other hand, MAPMT is more expensive as a device but could be directly coupled to a digitizer or other appropriate readout system.

The SiPM solution seems to be unavoidable for the detectors inside the spectrometer and the beam deflecting magnet. WLS fibers transmitting the light to the 16×16 multianode PMT Hamamatsu H9500 [9] might be a very suitable choice for the high energy positron veto due to its position – outside the magnetic field. The quantum efficiency of H9500 is ≈24% and the sensitive photocathode area per single pixel is 3.22×3.22mm².

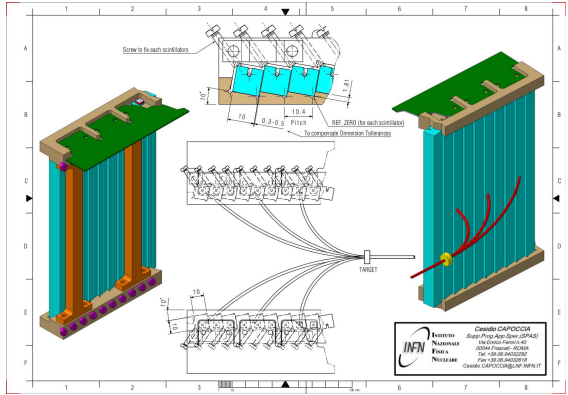

Figure 8. Possible design of the charged particle vetoes with scintillating strips rotated by 100 mrad

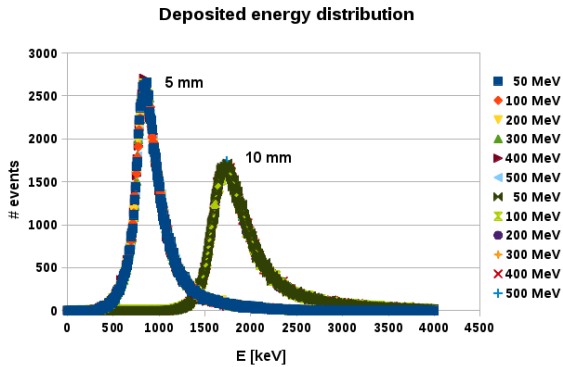

Figure 9. Deposited energy distribution for positrons impinging on a plastic scintillator. The first six series are for scintillator thicknesses of 5 mm while the last six – for 10 mm

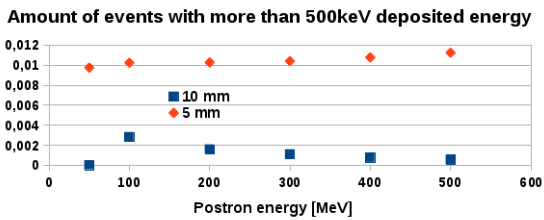

Figure 10. Inefficiency of scintillators with thicknesses of 5 mm and 10 mm and a minimal detectable energy of 500 keV

5. CONCLUSIONS AND DISCUSSIONS

While the realization of a hybrid fiber tracker with a CCD readout is state-of-the-art, the time scale of the PADME experiment does not allow spending much time on research and development of new detector technologies. That is why the most probable solution for the charged particle detectors will be based on extruded plastic scintillator bars, read out through WLS fibers either by SiPMs or multianode PMT. The fiber tracker design could still be very suitable for a possible future upgrade of the PADME experiment, when the main focus will be the detection of dark photon decays through the reaction A'→e⁺e⁻ and the position and the momentum resolution for the charged particles will be crucial for the background suppression.

***Acknowledgments.*** *Authors acknowledge support from University of Sofia science fund under grants FNI-SU N57/2015 and FNI-SU N39/2016, INFN, LNF-INFN under contract LNF-SU 70-06-497/07-10-2014, and Theta Consult Ltd. The authors would like to thank the colleagues from the Astronomy Department at the Faculty of Physics, University of Sofia for the possibility to use the SBIG 11000M CCD camera and Dr. Yavor Shopov for the focal length matching optical system.*